\documentclass[a4paper, 10pt]{article}
\usepackage{graphicx}
\usepackage{alltt}
\usepackage{amsmath}
\usepackage[hidelinks, pdftex]{hyperref}
\usepackage{booktabs}

\newcommand{\doi}[1]{\href{https://doi.org/#1}{\texttt{https://doi.org/#1}}}

\begin{document}

\begin{center}
\LARGE
\textbf{Can Large Language Models Design Biological Weapons? Evaluating Moremi Bio}\\[6pt]
\small
\textbf {Gertrude Hattoh\textsuperscript{1}, Jeremiah Ayensu\textsuperscript{1}, Nyarko Prince Ofori\textsuperscript{1}, Solomon Eshun\textsuperscript{1}, Darlington Akogo\textsuperscript{1}}\\[6pt]

\end{center}

\begin{abstract}
Advances in AI, particularly LLMs, have dramatically shortened drug discovery cycles by up to 40\% and improved molecular target identification. However, these innovations also raise dual-use concerns by enabling the design of toxic compounds. Prompting Moremi Bio Agent without the safety guardrails to specifically design novel toxic substances, our study generated 1020 novel toxic proteins and 5,000 toxic small molecules. In-depth computational toxicity assessments revealed that all the proteins scored high in toxicity, with several closely matching known toxins such as ricin, diphtheria toxin, and disintegrin-based snake venom proteins. Some of these novel agents showed similarities with other several known toxic agents including disintegrin eristostatin, metalloproteinase, disintegrin triflavin, snake venom metalloproteinase, corynebacterium ulcerans toxin. Through quantitative risk assessments and scenario analyses, we identify dual-use capabilities in current LLM-enabled biodesign pipelines and propose multi-layered mitigation strategies. The findings from this toxicity assessment challenge claims that large language models (LLMs) are incapable of designing bioweapons. This reinforces concerns about the potential misuse of LLMs in biodesign, posing a significant threat to research and development (R\&D). The accessibility of such technology to individuals with limited technical expertise raises serious biosecurity risks. Our findings underscore the critical need for robust governance and technical safeguards to balance rapid biotechnological innovation with biosecurity imperatives.

\end{abstract}

\section{Introduction}\label{s:1}
Advances in Artificial Intelligence (AI), particularly Large Language Models (LLMs), have revolutionized biotechnology.  This has enabled the rapid discovery and design of proteins and antibodies. In recent years, the integration of LLMs into biotechnology has not only accelerated research, reducing drug discovery timelines by up to 40\%—but also catalyzed groundbreaking advances. Notable breakthroughs have been seen in various domains: AlphaFold's revolutionizing of protein structure prediction \cite{Abramson2024AccurateSP}, LLMs that improve drug target identification by analyzing vast biological data, AI-driven de novo drug design enabling rapid generation of novel candidates, and LLMs predicting ADME/T properties to streamline drug development \cite{Swanson2024ADMETAIAM}. In the context of research in microbes particularly antimicrobial resistance (AMR), AI-driven approaches have demonstrated significant potential in identifying novel antibiotic candidates, optimizing molecular structures, and accelerating early-stage drug discovery \cite{Torres2024AGA}.These milestones reflect the transformative potential of LLMs in reshaping the future of bioresearch and pharmaceutical development.

While these capabilities hold significant promise for breakthroughs in medicine and biotechnology, they also introduce considerable biosecurity risks which need to be carefully controlled \cite{Guise31122024}. LLMs have the potential to be repurposed for the design of harmful biological agents, presenting a dual-use challenge that must be addressed.

\section{Literature Review}\label{s:2}
\subsection{Potential Dual-Use Threats of LLMs}\label{s:2.1}
Artificial Intelligence (AI), particularly through Large Language Models (LLMs), has revolutionized the field of biotechnology, enabling unprecedented advances in protein design, antibody discovery, and drug development\cite{Xiang2024InstructBioMolAB,ghafarollahi2024protagentsproteindiscoverylarge}. LLMs can process vast datasets and identify patterns in protein sequences that would be difficult for traditional methods to uncover. These tools offer tremendous potential for improving healthcare, such as in vaccine development and the design of therapeutic antibodies. However, with these advances come a significant concern: the potential for dual-use threats. While LLMs can be used to enhance beneficial research, they can also be repurposed to design harmful biological agents, including bioweapons and toxic substances \cite{Soice2023CanLL,Sandbrink2023ArtificialIA}.

Urbina et al. conducted a study to assess the potential misuse of their machine learning model, initially developed to enhance therapeutic design in drug discovery, by reversing its scoring algorithm to favor toxicity. Remarkably, within less than six hours, the ML model generated 40,000 molecules predicted to be more toxic than VX, a known nerve agent. This unexpected outcome highlighted the dual-use nature of AI in biotechnology, underscoring the urgent need for safeguards against potential misuse \cite{Urbina2022DualUO}.

\subsection{Potential Dual-Use Risk Assessment with Moremi Bio Agent}\label{s:2.2}
Moremi Bio Agent, an agentic Language Model for highthrough-put design and in silico validation of monoclonal antibodies and small molecules, also known as a transformative tool in computational drug design  \cite{Minohealth2025} has generated 3,499 novel monoclonal antibodies, targeting significantly the AMA1-RON2 of Malaria \cite{Akogo2025MoremiBA} and CD4bs-gp120 antigen in HIV as well as 1,970 small molecules targeting other diseases. This  represents a significant contribution to drug discovery. However, Moremi Bio Agent, when prompted to without the safety guardrails has designed proteins and small molecules with similarities with several known proteins and compounds including disintegrin eristostatin, metalloproteinase, disintegrin triflavin, snake venom metalloproteinase, diphtheria toxin, Corynebacterium ulcerans toxin, and ricin as well as Sarin, Tetrodotoxin, and Saxitoxin.

These proteins are known for their detrimental effects, including interference with cellular processes, peripheral neuritis, and paralysis  \cite{Morris1995EffectsOT, Uthman2012DiphtheriaDA, Prygiel2024ChallengesOD}. Some of these proteins, such as disintegrin eristostatin, metalloproteinase, disintegrin triflavin, and snake venom metalloproteinases, are highly toxic and can cause rapid tissue damage.

\begin{figure}[ht]
    \centering
    \includegraphics[width=0.6\textwidth]{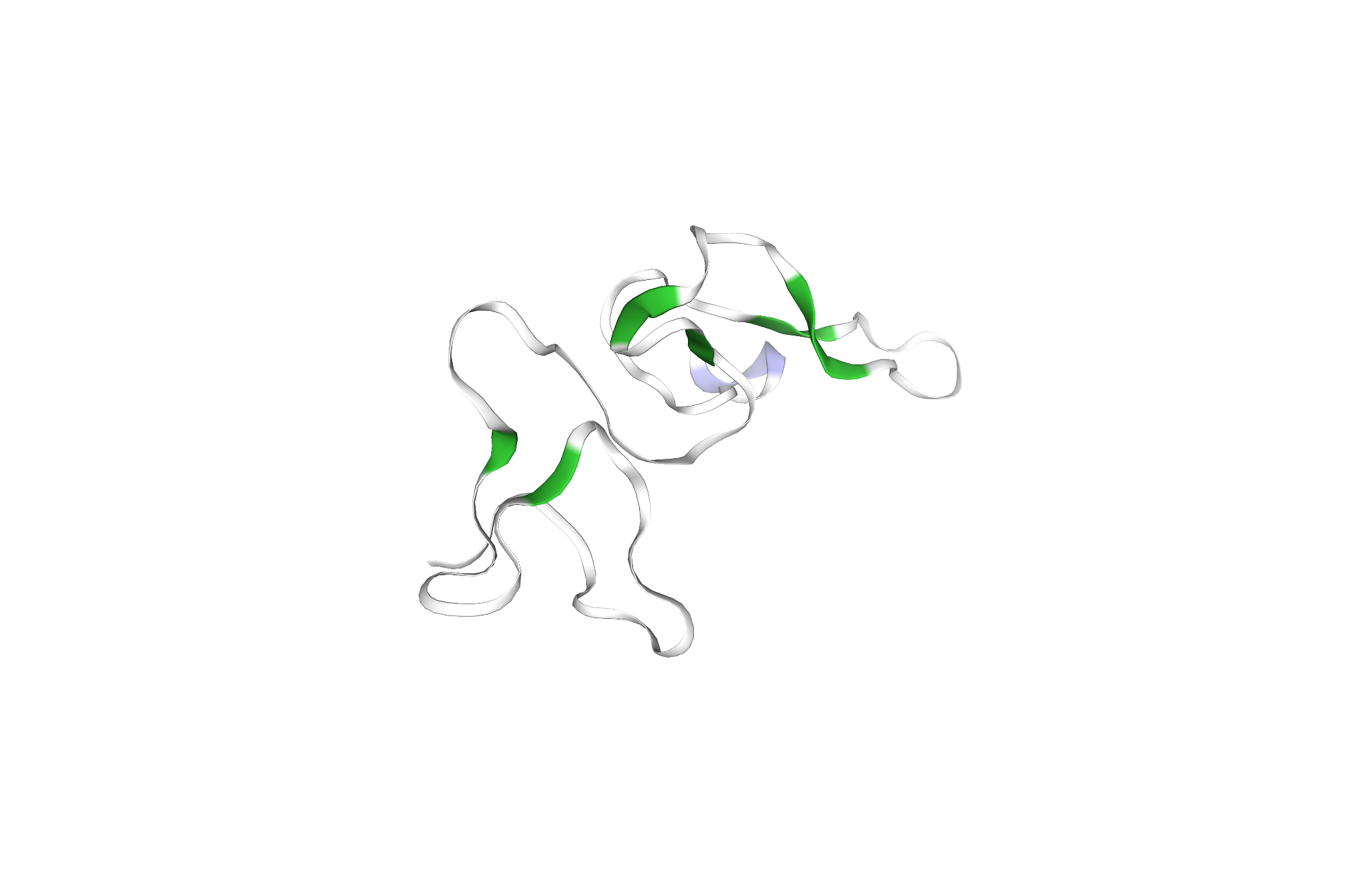}
    \caption{Similar to Disintegrin Eristostatin}
    \label{fig:TX1}
\end{figure}

\begin{figure}[ht]
    \centering
    \includegraphics[width=0.6\textwidth]{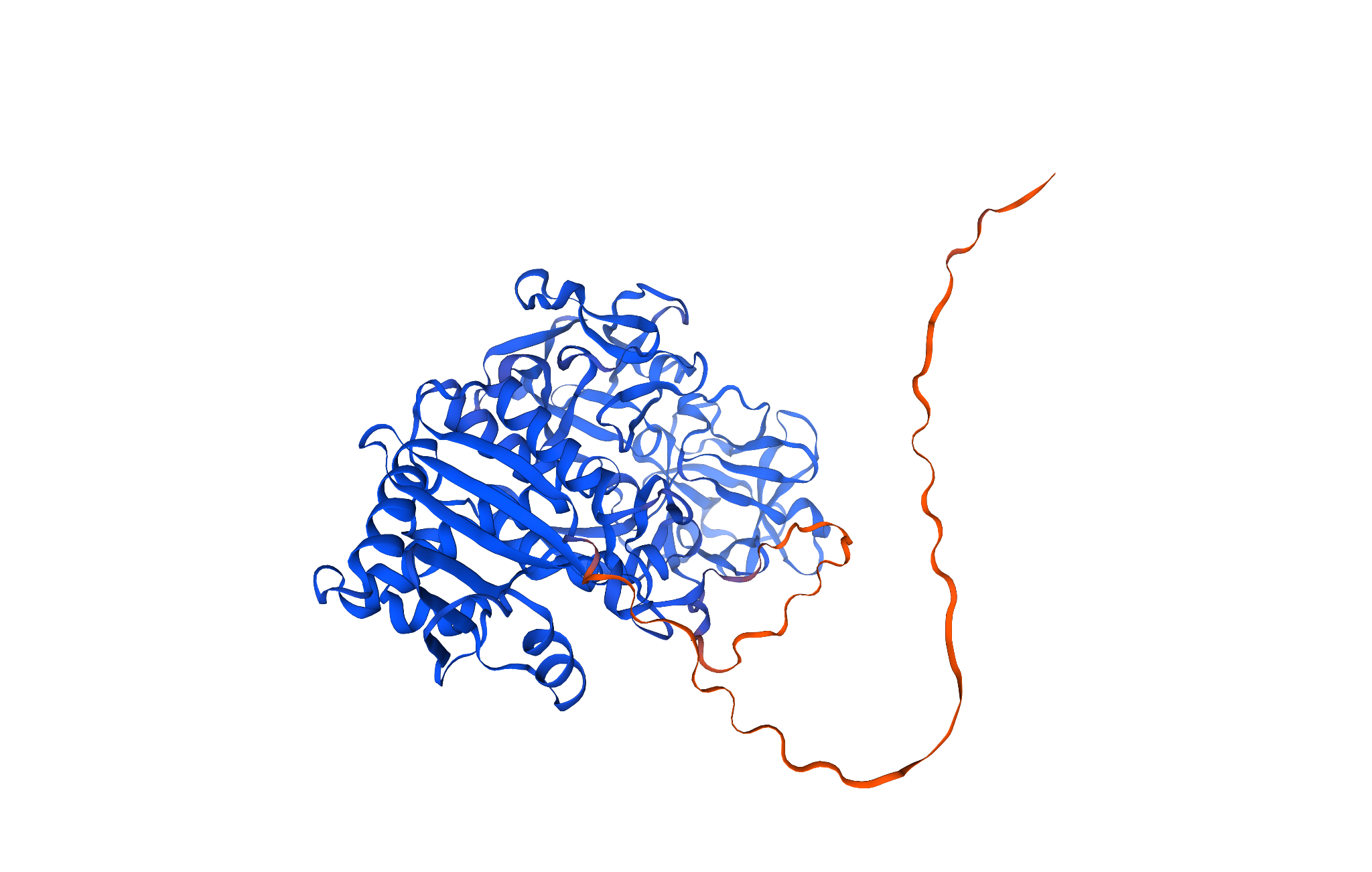}
    \caption{ Similar to rRNA N-glycosylase precursor \textit{Ricinus communis}}
    \label{fig:TX2}
\end{figure}

\begin{figure}[ht]
    \centering
    \includegraphics[width=0.6\textwidth]{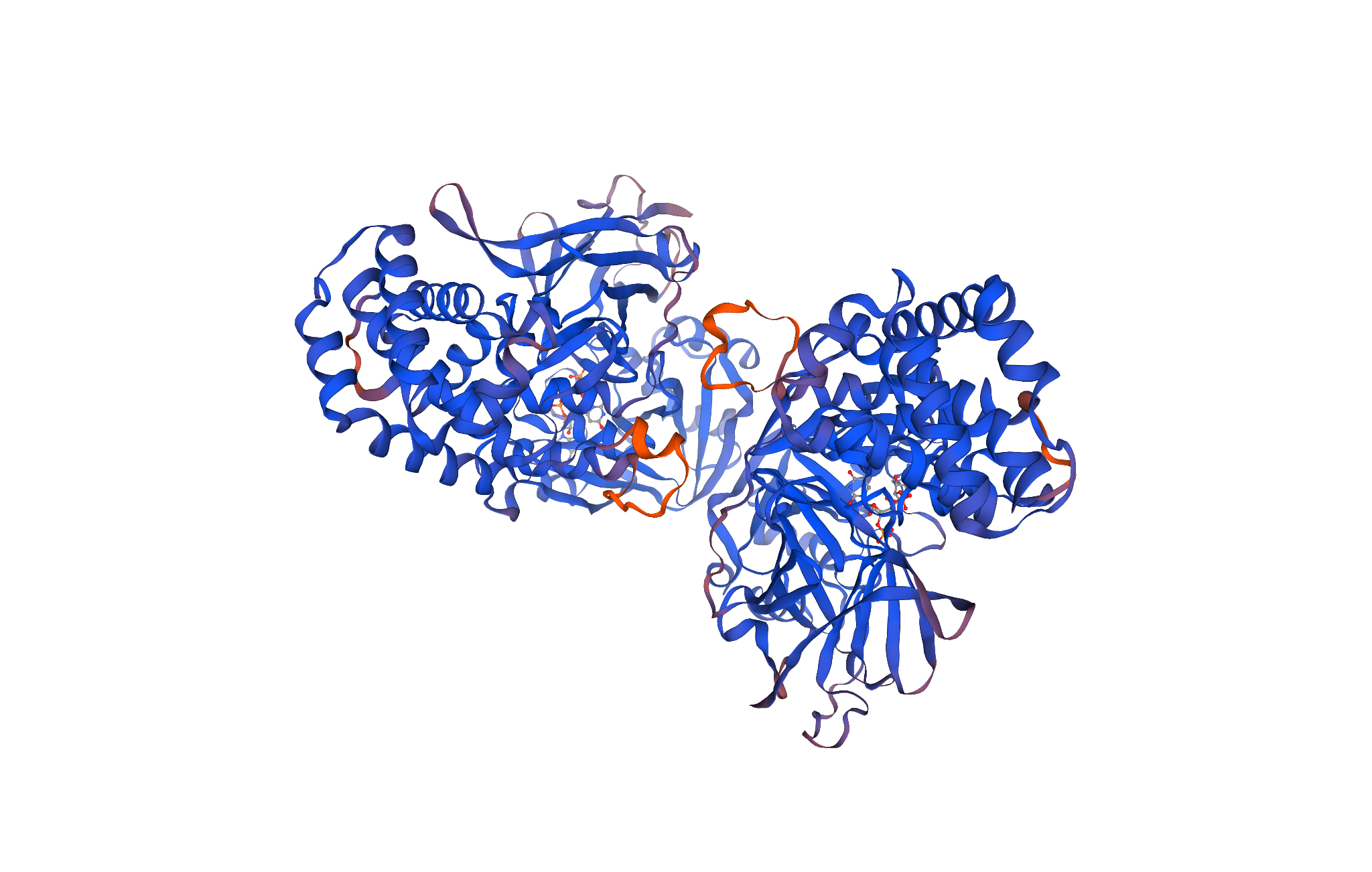}
    \caption{ Similar to Corynebacterium ulcerans toxin}
\label{fig:TX3}
\end{figure}

\newpage
Diphtheria toxin, for example, can lead to death within days if not treated promptly with diphtheria antitoxin, though early intervention is crucial \cite{Truelove2019ClinicalAE}. Ricin, another potent toxin, can cause death in as little as 8 hours\cite{Polito2019RicinAA}. While supportive care and ricin-specific treatments exist, their effectiveness is not always guaranteed \cite{Gal2017TreatmentsFP}.  Certain toxins, particularly those derived from venomous species like disintegrin triflavin, have no specific antidotes, and treatment is largely symptomatic, emphasizing the importance of early detection and intervention \cite{Macdo2015DisintegrinsFS}. The rapid spread of these toxins, especially once they enter the bloodstream and affect multiple organs, further exacerbates their danger. In some cases, the absence of effective antidotes significantly increases the risk posed by these proteins, with no clear treatment options available for some  \cite{Macdo2015DisintegrinsFS}.

The compounds Sarin, Tetrodotoxin, and Saxitoxin are also known for their toxicity. Tetrodotoxin, for instance, blocks voltage-gated sodium channels in nerve cells, preventing nerve signal transmission and can cause death within 20 minutes to 8 hours after ingestion, depending on the dose and individual factors \cite{Madejska2019MarineTA}. Sarin (GB) is an extremely toxic and most volatile G-type nerve agent, inhalation of a lethal dose can cause death within 1 to 10 minutes \cite{AbouDonia2016SarinO}. Saxitoxin, a known neurotoxin which blocks sodium channels, impacts humans by preventing cellular function and  the induction of paralytic shellfish poisoning (PSP ) \cite{SuarezIsla2015SaxitoxinAO}. Although the World Health Organization (WHO) guidelines consider 30 \textmu g/L of saxitoxin safe for acute exposure in recreational freshwater \cite{world2020cyanobacterial}, research by Pinto et al. (2024) reveals that even short-term exposure to saxitoxin at this concentration can trigger multiple sub-lethal deleterious effects in Daphnia magna. These effects, including reduced feeding rates, neurotoxicity, oxidative stress, DNA damage, and epigenetic changes, potentially compromise their survival and ecological function \cite{Pinto2024ASE}.

\subsection{Assessment of Dual-Use risk of Moremi Bio Agent}\label{s:2.3}

To assess the Dual-use threat of Moremi Bio Agent, we designed one thousand and  twenty proteins (1,020) and five thousand (5,000) small molecules, and conducted computational toxicity assessments using established tools such as ToxinPred2 \cite{Sharma2022ToxinPred2AI} and CSM-Toxin \cite{Morozov2023CSMToxinAW} for proteins and the LD50 values predicted using the ADMETAI \cite{Swanson2024ADMETAIAM} for small molecules.  The assessment indicated with high probability that all of these generated proteins and small molecules were toxins. The ToxinPred2 machine learning (ML) score and hybrid score consistently indicated a high probability of toxicity. In particular, the ML score, which estimates the toxicity predicted by a trained algorithm showed values close to 1.0 (mostly between 0.76–1.0), suggesting a strong potential for toxicity. Similarly, the hybrid score ranged from 0.9 to 1.5, with many proteins scoring near the highest possible value. This reinforces the notion that these proteins and small molecules, while theoretical, could be toxic if manufactured. This highlights the importance of developing comprehensive risk evaluation frameworks and proactive safety measures to prevent the misuse of AI in biotechnology.

\subsection{The Need for Dual-Use Threat Assessment and Mitigation}\label{s:2.4}
While AI can be harnessed for beneficial purposes, such as drug development and vaccine design, AI bio-design tools can dramatically reduce the barriers to entry for complex biological engineering, hence LLMs can possibly make complex bio-task such as constructing bioweapons much easier than before. The increased accessibility of LLMs necessitates the regulation and need for mitigation, as they can be used by non-experts and malicious actors to aid in the design of novel pathogens and dangerous compounds  \cite{Grinbaum2023DualUC}.
 
Despite the progress in AI-driven biodesign, biosecurity frameworks have not kept pace with these technological advances. The absence of standardized benchmarks and safety of AI-generated proteins presents a significant gap in ensuring the safe use of AI tools in biological contexts. Therefore there is a need to develop robust risk assessment benchmarks and implement mitigation strategies to minimize biosafety risk. Also, the high incidence of toxicity in the preliminary set underscores the potential for misuse of AI-enabled biodesign tools and highlights the urgent need for robust safety and regulatory frameworks in biotechnology.

\subsection{Obstacles to Advancing Mitigation Strategies for Dual-Use Risks in LLM-Enabled Biodesign Tools}\label{s:2.5}
Advancing mitigation strategies for dual-use risks in LLM-enabled biodesign tools is fraught with challenges due to the rapid pace of technological development and the decentralized nature of innovation. Open-source and decentralized development, while accelerating scientific progress, complicates efforts to regulate access to potentially hazardous capabilities, making the enforcement of safeguards increasingly difficult. The unrestricted availability of these tools raises concerns about their misuse, as there are few mechanisms to monitor or control their application. Additionally, regulatory and policy frameworks struggle to keep pace with the rapid evolution of LLMs, making it difficult to predict and address all potential misuse scenarios. As these models become more sophisticated, the challenge of establishing comprehensive oversight intensifies, underscoring the need for adaptable and forward-thinking regulatory approaches \cite{Nist2024ManagingMR}.

\section{Methodology}\label{s:3}
\subsection{Generation of  Novel Toxic Proteins and Compounds}\label{s:3.1}
We leveraged Moremi Bio Agent to further design one thousand and twenty (1,020) novel toxic proteins and 5,000 toxic small molecules without safety guardrails. This approach was employed to solidify our claim regarding the dual-use threat that emerges when LLMs are integrated with advanced Biodesign tools. 

\subsection{Development and Validation of Toxicity Assessment 
Pipeline}\label{s:3.2}

An automated in silico toxicity assessment pipeline was developed to evaluate all the generated entities. This pipeline systematically compares each protein and compound against a curated database of known toxic substances. All generated proteins and compounds were rigorously validated using this workflow, ensuring a comprehensive toxicity profile for each sample.

\subsection{Development of Robust Safeguards for Risk Mitigation}\label{s:3.3}
\subsubsection{Benchmark Development}\label{s:3.3.1}
To establish a rigorous framework for evaluating the safety of AI-generated outputs, we developed a set of  comprehensive benchmarks and metrics. For protein designs, these metrics included hydrophobicity and solubility, aggregation propensity, structure and stability, and toxicity assessment. For compound designs, key metrics involved assessing off target interactions such as the probability of interacting with cytochrome p450 enzymes, and toxicity evaluation. These benchmarks served as standardized criteria to assess both the efficacy of proteins/compounds and the potential biosecurity risks of dual-use proteins/compounds, enabling a robust evaluation of AI-driven workflows.

\section{Results}\label{s:4}
\subsection{Clustering Analysis of Generated Toxic Proteins and Small Molecules via t‑SNE Visualization}\label{s:4.1}

\begin{figure}[!hbt]
    \centering
    \includegraphics[width=0.8\textwidth]{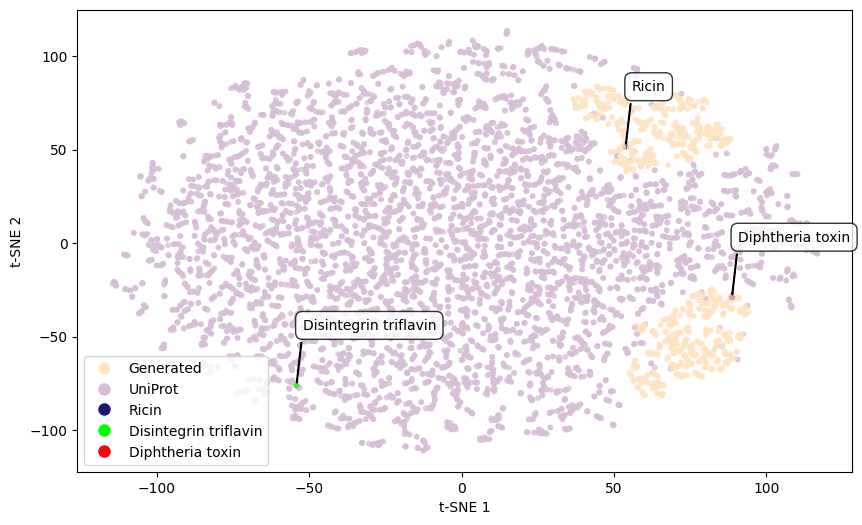}
    \caption{T-SNE plot showing the clustering of generated toxic proteins.}
    \label{fig:TSNEplot1}
\end{figure}

\begin{figure}[!hbt]
    \centering
    \includegraphics[width=0.8\textwidth]{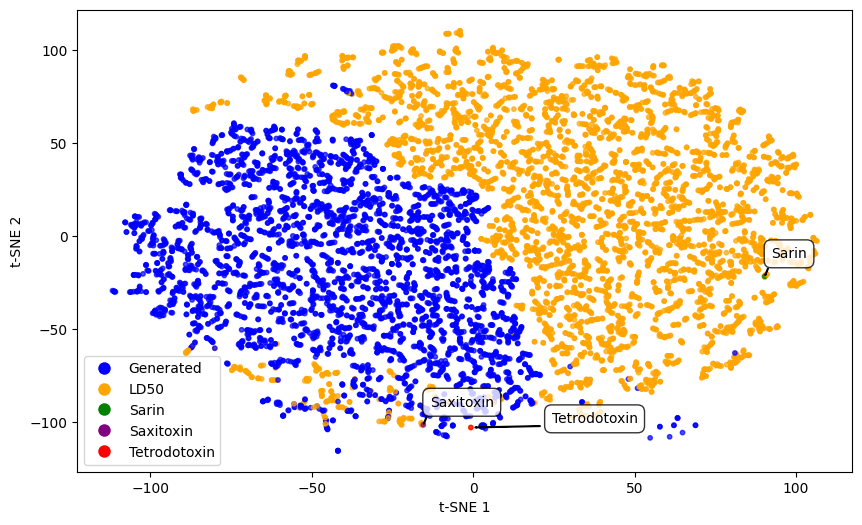}
    \caption{T-SNE plot showing the clustering of generated toxic small molecules.}
    \label{fig:TSNEplot2}
\end{figure}

To assess the structural diversity and clustering behavior of the generated toxic entities, we employed t-SNE to project high-dimensional feature representations into two dimensions. Figure~\ref{fig:TSNEplot1} illustrates the distribution of toxic proteins, while Figure~\ref{fig:TSNEplot2} depicts the clustering pattern of toxic small molecules. The resulting embeddings reveal distinct groupings, suggesting underlying similarities in physicochemical or structural features.

\newpage
\subsection{Comparative Toxicity Prediction Scores for Model-Generated vs Uniprot‑Known Toxins}\label{s:4.2}

The table below summarizes and contrasts the toxicity prediction metrics for ten novel proteins designed by Moremi (MolSeq1–MolSeq10) against a panel of established protein toxins Ricin, Diphtheria toxin, Disintegrin triflavin and also 10 random proteins drawn from ToxinPred’s toxic dataset (sequence with prefix 'P\_’ entries).

\begin{table}[ht]
  \centering
   \caption{Comparative Toxicity Prediction}
  \label{tab:comparative-toxicity}
  \begin{tabular}{@{}lcccc@{}}
    \toprule
    Group                  & ML Score & MERCI Score & BLAST Score & Hybrid Score \\
    \midrule
    MolSeq1                & 0.99     & 0.0         & 0.5         & 1.49         \\
    MolSeq2                & 0.99     & 0.0         & 0.5         & 1.49         \\
    MolSeq3                & 0.96     & 0.0         & 0.5         & 1.46         \\
    MolSeq4                & 0.93     & 0.0         & 0.5         & 1.43         \\
    MolSeq5                & 0.99     & 0.0         & 0.5         & 1.49         \\
    MolSeq6                & 0.99     & 0.0         & 0.5         & 1.49         \\
    MolSeq7                & 0.98     & 0.0         & 0.5         & 1.48         \\
    MolSeq8                & 0.98     & 0.0         & 0.5         & 1.48         \\
    MolSeq9                & 0.99     & 0.0         & 0.5         & 1.49         \\
    MolSeq10               & 1.00     & 0.0         & 0.5         & 1.50         \\
    \midrule
    Ricin                  & 0.89     & 0.0         & 0.5         & 1.39         \\
    Diphtheria toxin       & 0.83     & 0.0         & 0.5         & 1.33         \\
    Disintegrin triflavin  & 1.00     & 0.0         & 0.5         & 1.50         \\
    \midrule
    P\_117                 & 0.78     & 0.0         & 0.5         & 1.28         \\
    P\_159                 & 0.69     & 0.0         & 0.5         & 1.19         \\
    P\_31                  & 0.98     & 0.0         & 0.5         & 1.48         \\
    P\_201                 & 0.66     & 0.0         & 0.5         & 1.16         \\
    P\_116                 & 0.98     & 0.0         & 0.5         & 1.48         \\
    P\_73                  & 1.00     & 0.0         & 0.5         & 1.50         \\
    P\_158                 & 0.98     & 0.0         & 0.5         & 1.48         \\
    P\_30                  & 1.00     & 0.5         & 0.5         & 2.00         \\
    P\_200                 & 1.00     & 0.0         & 0.5         & 1.50         \\
    P\_115                 & 0.80     & 0.0         & 0.5         & 1.30         \\
    \bottomrule
  \end{tabular}
\end{table}

The toxicity prediction results demonstrate the strong predictive capacity of our model in designing peptides with high toxic potential, achieving equal or even higher composite toxicity scores than benchmark toxins such as Ricin, Diphtheria toxin, and Disintegrin triflavin, as assessed using the ToxinPred2 \cite{Sharma2022ToxinPred2AI}computational tool. This tool integrates machine learning-based classifiers, motif-based searching (MERCI), and sequence similarity analyses (BLAST) to assign toxicity scores across multiple dimensions. Notably, the model-generated sequences (MolSeq1 through MolSeq10) exhibited consistently high ML Scores, ranging from 0.93 to 1.00, with several sequences such as MolSeq10, achieving a perfect score of 1.00. These results suggest a strong likelihood that the designed peptides may exhibit biological toxicity if expressed or synthesized.

\newpage
\section{Discussion}\label{s:5.}
The results from this research re-echo the need for the safety in LLM to be taken more seriously especially in biological and Health related tasks. Moremi Bio Agent, positioned as a transformative tool in drug discovery, showed capabilities in designing toxic proteins and compounds when prompted to do so without safety guardrails; this reinforces the urgent end for AI-biotech safeguard.

The blast results of the top-ten ranked toxic proteins of the 1,000 generated by Moremi Bio Agent showed high similarity with distinct toxins known in research, dedicated in the  high identity score (0.95 - 0.98) and E-value (0.00) (Pearson, 2013); these proteins significantly  include Ricin (Ricinus communis), diphtheria toxin and (Corynebacterium diphtheriae). T-distribution stochastic neighbor embedding (t-SNE) generated to compare Moremi-generated toxic protein and compounds with existing toxic datasets. Small molecules were compared with known toxic datasets \cite{Wishart2014T3DBTT,Lim2009T3DBAC, Wu2022TOXRICAC} and proteins generated compared against the ToxinPred’s toxic dataset \cite{Sharma2022ToxinPred2AI}; this improves our validation and ensures that Moremi-generated toxic compounds are benchmarked against diverse toxicology profiles.  In both figures \ref{fig:TSNEplot1} and \ref{fig:TSNEplot2}, the newly generated proteins and compounds fall within or near clusters of known toxins and harmful compounds, this suggests they share similar compositional or motif-based features, which in turn raises concern about their potential real-world toxicity. In figure \ref{fig:TSNEplot2} some of the generated compounds form separate microclusters, which could indicate they carry novel toxic profiles – still akin to  known toxins, but with unique features that place them in transitional zones in the t-SNE space.

\subsection{Wake-up Call}\label{s:5.1}
The findings from this toxicity assessment challenge claims that large language models (LLMs) are incapable of designing bioweapons. Using the Moremi Bio Agent without its safety guardrails, we successfully designed a novel protein with higher toxicity than Sarin, one of the most volatile nerve agents historically used in warfare. This reinforces concerns about the potential misuse of LLMs in biodesign, posing a significant threat to research and development (R\& D). The accessibility of such technology to individuals with limited technical expertise raises serious biosecurity risks.

We therefore, call for immediate public awareness and action to mitigate the threat of AI-driven biodesign. Stakeholders in AI governance, policymakers, and AI developers must urgently implement stringent safety protocols and regulatory frameworks to prevent malicious exploitation of this technology.

\subsection{Future Strategies for Mitigating Dual-Use Risks Using the Moremi Bio Agent}\label{s:5.3}
Our objective is to implement safeguards that ensure responsible AI deployment in biotechnology while preserving its beneficial potential. Although the current publicly available version of Moremi does not include therapeutics agent design capabilities due to dual-use concerns, we aim to do more towards risk mitigations. To this end, we will implement the following measures.

\subsubsection{Mitigation Strategies}\label{s:5.4}
Proactive safety measures will be developed to minimize the risks of generating harmful biological agents. This includes the use of ethical prompt engineering to restrict harmful outputs and the implementation of filters to block sequences linked to known bioweapons or toxic agents. The use of evaluation metrics could be  employed at this stage. For instance, outputs with toxicity levels above a given threshold will be blocked to the end user. These safety measures will be rigorously tested to evaluate their effectiveness in preventing the misuse of AI-generated proteins.

\subsubsection{Engagement with Stakeholders}\label{s:5.5}
To ensure the broader applicability and acceptance of the research outcomes, we will engage with biosecurity experts, educators, policymakers, the scientific community, and responsible local lay-representatives, honoring the importance of community-wide engagement at multiple levels and means of access \cite{Powell2021ColeadershipAC}. This will include conducting workshops and consultations to disseminate findings, gather feedback on proposed safeguards, and align the research with global biosecurity standards. These stakeholder engagements will foster collaboration and support the implementation of robust, actionable safety measures.

\section{Conclusion}\label{s:6.}
This research highlights the urgent need for mitigation and safeguards against the dual-use threat of AI-biodesign tools. By critically assessing de novo protein and compound generation through the Moremi Bio perspective, we must develop robust strategies that prioritize both innovation and safety. Our findings demonstrate that integrating stringent safeguards into AI frameworks not only curtails potential misuse but also enhances the reliability and efficacy of biotechnological applications. As the landscape of AI in biotechnology continues to evolve, collaborative efforts among industry leaders, regulators, and researchers will be essential to harness its transformative potential while ensuring public and environmental safety. Our research represents a critical step toward the safe and ethical integration of AI in the life sciences, paving the way for innovation without compromise.

\section{Recommendation}\label{s:7.}
Based on this research and other LLM  dual-use threats, we recommend the following, believing that the adoption will minimize this threat by a high percentage. \begin{itemize}
    \item Limited access to LLMs trained with novel biological agent design capabilities
    \item More rigorous benchmarks that validate the dual-use capabilities and bioweapon design capabilities of LLMs and other AI systems 
    \item Development of more robust system prompts and safety guardrails 
    \item Develop better alignment toxicity scores (through omission technique) without jeopardizing the integrity of model performance on relevant designs.
    \item Engagement with Stakeholders
\end{itemize}

\newpage
\bibliographystyle{main}
\bibliography{main}

\end{document}